\documentclass[a4paper,12pt,oneside,english]{report}
\usepackage{natbib} 
\usepackage{mathptmx}
\usepackage{amsmath}
\usepackage{amsfonts}
\usepackage{amssymb}
\usepackage{rotating}
\usepackage{subfigure}

\usepackage{graphicx}
\usepackage{babel}
\usepackage{epsfig}

\newcommand{\aap}{    {\it Astron. Astrophys.}}

\newcommand{\apj}{    {\it Astrophys. J.}}
\newcommand{\apjl}{   {\it Astrophys. J. Lett.}}

\newcommand{\grl}{    {\it Geophys. Res. Lett.}}

\newcommand{\solphys}{{\it Solar Phys.}}

\makeatother
\begin{document}
%
%

{\Large{\bf{Solar Polar Fields During Cycles 21 --- 23: Correlation with Meridional Flows}}}\\

{\large{P.~{Janardhan}$^{1}$
        K.B.~{Susanta}$^{1}$
        S.~{Gosain}$^{1,2}$
       }}\\\\

${^{1}}$ Physical Research Laboratory, Astronomy \& Astrophysics Division, Navrangpura, 
   Ahmedabad - 380 009, India.\\
email: jerry@prl.res.in email: susanta@prl.res.in ~~ \\\\ 
${^{2}}$ Udaipur Solar Observatory, P.O. Box 198, Dewali, Udaipur 313001, India.\\
email: sgosain@prl.res.in ~~ \\\\\\\\
             
{\Large{\bf{Abstract}}}\\\\
We have examined polar magnetic fields for the last three solar cycles, {$\it{viz.}$}, cycles 
21, 22 and 23 using NSO Kitt Peak synoptic magnetograms. In addition, we have used SoHO/MDI 
magnetograms 
to derive the polar fields during cycle 23.  Both Kitt Peak and MDI data at high latitudes 
(78${^{\circ}}$--90${^{\circ}}$) in both solar hemispheres show a significant drop in the 
absolute value of polar fields from the late declining phase of the solar cycle 22 to the 
maximum of the solar cycle 23. We find that long term changes in the absolute value of the 
polar field, in cycle 23, is well correlated with changes in meridional flow speeds that 
have been reported recently.  We discuss the implication of this in influencing the extremely 
prolonged minimum experienced at the start of the current cycle 24 and in forecasting the 
behaviour of future solar cycles. 


\section*{Introduction}\label{S-Intro} 

Detailed studies of solar magnetic features is an important area of
research because solar magnetic fields have a profound and far reaching 
influence on the earth's near-space environment.  With mankind's increased 
dependance on space based technology, much of which is at risk due to 
solar activity that waxes and wanes with the sunspot cycle, 
it is imperative that we understand the solar magnetic cycle and 
its effects on the 
near-space environment. In addition, due to the significant anthropogenic 
influence on climate change in recent times, it is becoming increasingly important to 
distinguish and delineate the degree to which the solar cycle can affect terrestrial climate. 

The evolution of large scale solar magnetic fields is attributed to a solar magnetohydrodynamic 
dynamo operating inside the sun which involves three basic processes ${\it{viz.}}$ 
the generation of toroidal fields by shearing pre-existing poloidal fields (the 
$\Omega$ effect); the regeneration of poloidal fields by twisting toroidal flux 
tubes (the $\alpha$ effect); and finally, flux transport by meridional circulation 
to carry background magnetic fields poleward from the equator \cite{Par55a,Par55b,SKr69,WSN91,CSD95,DCh99}.  

With the measurement of the sun's polar field \cite{BBa55} and the subsequent  
proposal of polar field reversal during maximum of each solar cycle \cite{Bab59}, 
research in solar magnetic fields and its effect on subsequent cycles was channeled in a 
new direction. Since then many investigations have been carried out to explore the 
relation between evolution of large-scale magnetic fields and their association 
with polar field structures \cite{FMW98,BKS01,BKS02,GoL03}).  

The current sunspot minimum that we have seen at the end of cycle 23 have been 
one of the deepest minima that we have experienced in recent times with roughly 
71\%----73\% of the days in 2008 and 2009 respectively being entirely spotless.  
Apart from this, cycle 23 has shown several other peculiarities, like a second maximum
during the declining phase that is unusual of odd numbered cycles, a slower rise 
to maximum than other odd numbered cycles and a slower than average polar reversal.  
Such departures from ``${\it{normal}}$" behaviour need to be studied and understood as 
they could be significant in the context of understanding the evolution of magnetic 
fields on the sun. 

In this paper, we have examined the solar polar field behavior for the last few cycles and 
the following sections will discuss our findings.  

\section*{Magnetic Field Data}
\subsection*{Low resolution NSO/Kitt Peak data}

Magnetic field data is available as standard FITS format files 
from the National Solar Observatory, Kitt Peak, USA (NSO/Kitt Peak) synoptic magnetogram 
database (ftp://nsokp.nso.edu/kpvt/synoptic/mag/).  
These synoptic maps are in the form of 180$\times$360 arrays starting from Carrington Rotation 
(CR) number CR1625 through CR2006 corresponding to years 1975.13 through 2003.66. For the preriod covering the 
years from 2003.66 through 2009.11, synoptic maps were obtained from the Vector Spectro Magnetograph (VSM) 
of the NSO Synoptic Optical Long-term Investigations of the Sun (SOLIS) facility for solar observations 
over a long time frame (NSO/SOLIS) (ftp://solis.nso.edu/synoptic/level3/vsm/merged/carr-rot/). A 
few data gaps during CR1640-CR1644, CR1854, CR1890,CR2015, CR2016, CR2040 and CR2041 were filled in by 
interpolated values while making the plots.

The resolution of the synoptic maps is 1${^{\circ}}$ in both longitude (1${^{\circ}}$ to 360${^{\circ}}$) 
and latitude(-90${^{\circ}}$ to 90${^{\circ}}$).  A typical Carrington synoptic map is produced from 
full disk magnetograms spanning over an entire Carrington rotation period. Each individual magnetogram is 
first remapped into latitude and longitude coordinates and then added together to produce the final synoptic 
map for each CR. All synoptic magnetogram maps, represented as an i $\times$ j data array 
with i and j representing the latitude and longitude respectively, for any given CR number n, contain 
magnetic flux density $\phi_ {i,j,n}$, in units of Gauss averaged over equal 
areas of the sun. The actual flux units are then obtained by multiplying by the appropriate area. 

Since we are interested in the variation of photospheric magnetic fields with latitude, 
data is processed by taking longitudinal averages for each latitude bin of 1${^{\circ}}$ by converting the data to a one dimensional array of 1 $\times$ 180.  Any latitude element i, for a given CR, n will contain the averaged magnetic field represented by equation (\ref{eq1}).
\begin{equation}
 \mathsf{ \phi_{i,n} = \frac{\sum_{{j=1}}^{360} \phi_{i,j,n}}{360}}
\label{eq1}
\end{equation}
Now, the averaged magnetic field for a range of latitudes in any CR is obtained 
by averaging $\phi_{i,n}$ as shown by equation (\ref{eq2}).
\begin{equation}
 \mathsf{ \phi_{n} = \frac{\sum_{{i=k}}^{p} \phi_{i,n}}{p-k}}
 \label{eq2}
\end{equation}
Where k and p are the row numbers corresponding to a latitude bin. Using the above procedure, we have obtained average magnetic fields in each solar hemisphere for three different latitude bins between ranges
\begin{itemize}
\item 0${^{\circ}}$ and 45${^{\circ}}$ -- representing the equatorial or toroidal fields;
\item 45${^{\circ}}$ and 78${^{\circ}}$ -- representing the mid latitude fields; 
\item and 78${^{\circ}}$ and 90${^{\circ}}$ -- representing polar fields in each hemisphere. 
\end{itemize}

\subsection*{High resolution MDI Data}

Line-of-sight magnetograms from the Michelson Doppler Interferometer (MDI, \cite{ScB95}) 
onboard the Solar and Heliospheric Observatory (SoHO; \cite{DFP95}), are available from 1996 
onwards.  High resolution (3600$\times$1080 pixels) MDI synoptic images beginning from 
CR1911 through CR2080 are used, with data gaps for CR1938-CR1942, CR 1945 and CR2073. 
It must however be noted that there are occasions when the inclination of the earth's orbit to the 
helioequator will cause one of the poles of the sun to not be clearly visible.  We have used 
MDI synoptic magnetogram available at http://soi.stanford.edu/magnetic/index6.html that have 
been corrected for this effect.  

Since MDI measurements are considered to be more reliable 
than Kitt-Peak observations, we have compared the two data sets for the period after 1996 to check 
if the results agree with each other.  Also, data gaps when present, for 
example from CR1938 to CR1941 were dealt with by replacing them with interpolated 
values, and these have been indicated in the figures. Finally, the resolution of the MDI data was degraded 
(by averaging) to the resolution of the NSO/Kitt Peak data so as to be able to compare the results. It 
may be noted that for the MDI data, only the latitude bin between  78${^{\circ}}$ and 90${^{\circ}}$ was considered.
%
\begin{figure}[ht]
\begin{center}
\includegraphics[width=0.55\textwidth,angle=0]{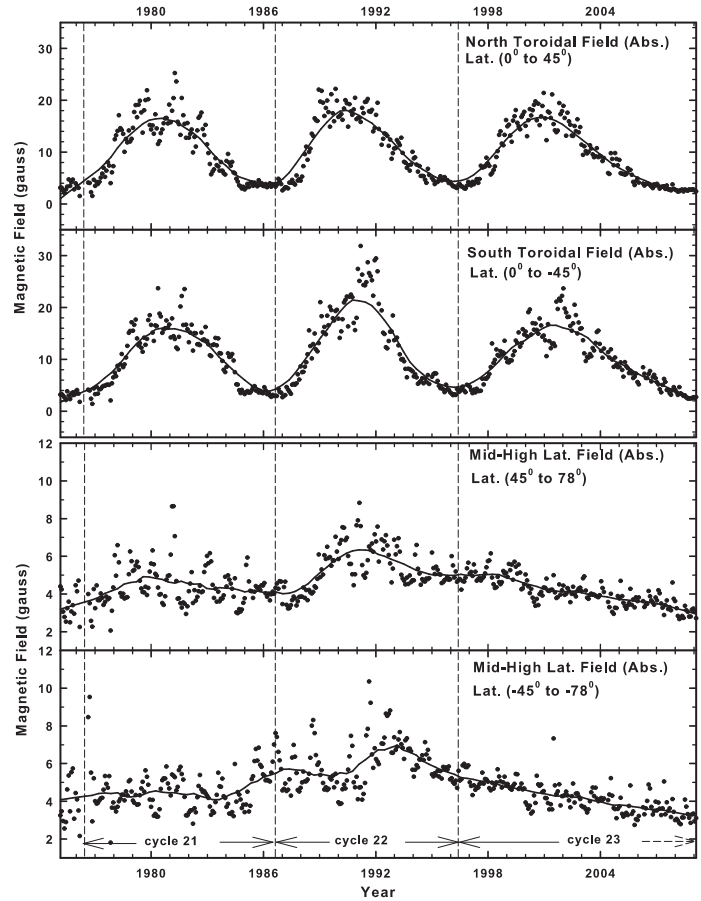}
\end{center}
\caption{{\sl\small{The four panels, in pairs starting from the top, show the variations in 
the magnetic field in the north and south solar hemisphere respectively, as a function of years 
for the solar cycles 21--23.  The absolute value of the magnetic field in the latitude range 
0${^{\circ}}$--$\pm45{^{\circ}}$ is shown in the first pair of panels while the second pair of 
panels show the absolute value of magnetic field in the latitude range 45${^{\circ}}$--78${^{\circ}}$. 
The filled dots represent the actual measurements while the solid line is a smoothed curve. 
The vertically oriented dashed parallel lines demarcate cycles 21, 22 and 23 respectively.}}}
\label{torr-fields}
\end{figure}	    
%

\section*{Magnetic Field Measurements}\label{S-Obs}

As mentioned earlier, meridional circulation is the primary poloidal flux transport 
agent in the sun.  A poleward meridional flow at the solar surface has been
shown to exist, with average speeds in the range 10 m s${^{-1}}$~--~20 m s${^{-1}}$ \cite{Duv79,Hat96,HaG96}. 
Since the amplitude of this surface flow is more than an order 
of magnitude weaker than other surface 
flows like granulation, supergranulation and differential rotation, the effects of this meridional 
circulation can be best studied at high solar latitudes where the modulation due to the solar cycle is 
negligable or absent.  
Figure \ref{torr-fields} shows the measurements of the magnetic field in each solar hemisphere 
as a function of time in years in the two latitude ranges 0${^{\circ}}$--45${^{\circ}}$ and 
$45{^{\circ}}$--$78{^{\circ}}$. While the upper two panels of Fig. \ref{torr-fields} 
show the absolute value of the field in the latitude range 0${^{\circ}}$--45${^{\circ}}$ 
for the northern and southern hemisphere respectively, the lower two panels show the 
absolute value of the magnetic field in the latitude range 45${^{\circ}}$--78${^{\circ}}$ for the 
northern and southern hemisphere respectively.  The filled dots represent the actual measurements  
derived from the NSO/Kitt Peak data base, while the solid line is a smoothed curve representing the data.
The vertically oriented dashed parallel lines demarcate cycles 21, 22 and 23.  It can be seen from Fig. \ref{torr-fields} (upper two panels) that there is a 
strong solar cycle modulation present in the latitude range $\pm0{^{\circ}}$--$\pm45{^{\circ}}$ while 
the solar cycle modulation is much weaker and barely descernable in the latitude range $\pm45{^{\circ}}$--$\pm78{^{\circ}}$ for cycles 21 and 22 while it is not seen for cycle 23.
%
\begin{figure}[ht]
\begin{center}
\includegraphics[width=0.55\textwidth,angle=0]{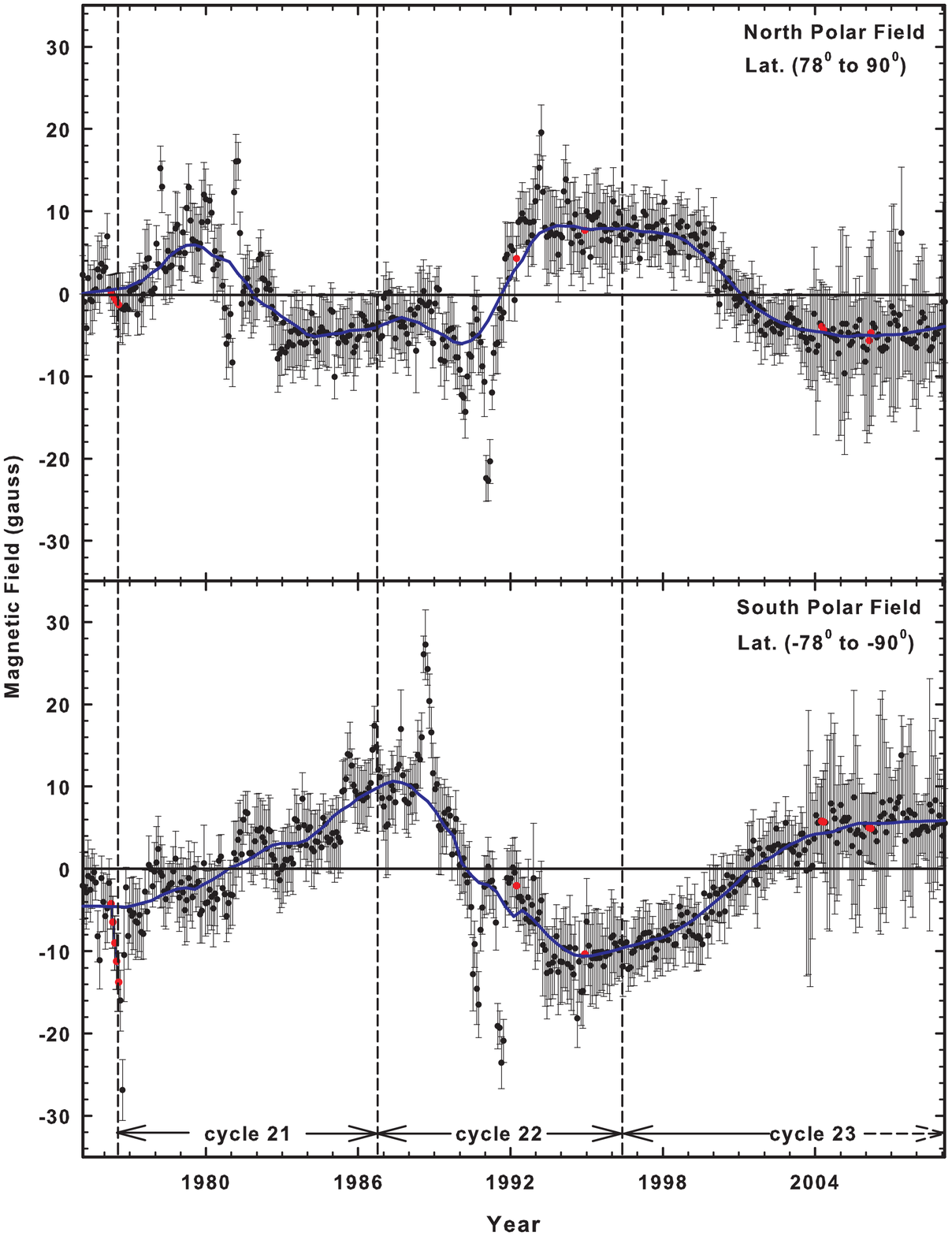}
\end{center}
\caption{{\sl\small{The panels show the variations in the magnetic field in the north (upper panel) 
and south solar hemisphere (lower panel) as a function of years for the solar cycles 21--23.  
The actual (signed) magnetic field in the latitude range 78${^{\circ}}$--90${^{\circ}}$ is shown 
in the two panels. The filled dots, with 1$\sigma$ error bars represent the actual measurements 
while the blue solid line is a smoothed curve. The vertically oriented dashed parallel lines 
demarcate cycles 21, 22 and 23 respectively. Points where data gaps were filled by interpolation are 
shown by red dots.}}}
\label{polar-fields}
\end{figure}	    
%

Figure \ref{polar-fields} shows the actual (signed) measurements of magnetic field, in each solar 
hemisphere, as a function of time in years in the latitude range  $78{^{\circ}}$--$90{^{\circ}}$. 
While the upper panel of Fig. \ref{polar-fields} show the actual (signed) value of the field 
for the northern hemisphere, the lower panel shows the actual (signed) value of the magnetic field  
for the southern hemisphere.  The filled dots, with 1 $\sigma$ error bars, represent the measurements 
derived from the NSO/Kitt Peak data, while the solid blue line is a smoothed curve.
Points where data gaps were filled by interpolation are shown by red dots. The vertically oriented 
dashed parallel lines demarcate cycles 21, 22 and 23.  A comparison of the the polar 
field during cycles 22 and 23 from Fig. \ref{polar-fields} shows that the longitude averaged polar magnetic field
between  $\pm78{^{\circ}}$--$\pm90{^{\circ}}$ during the current extended minimum leading up to the 
current cycle 24 is weaker than the corresponding minimum period in cycles 21 and 22.  Also, a 
comparison of the fields in the two hemispheres, in cycle 23, shows that the south polar field is 
weaker than north polar field.  In a detailed analysis of sunspot data and longitude-averaged photospheric
magnetic flux \cite{DiG04} has als pointed out several peculiarities during cycle 23 like the  
slow build up of the polar field after the occurence of the polar field reversal, the 
asymmetry in polar reversal for north and south pole and the steady behavior of the north polar field
in comparison to south polar field from the late declining phase of cycle 22 to early rising phase 
of cycle 23.  All of these features can also be clearly seen in Fig. \ref{polar-fields} for the period 
of cycle 23. 

\begin{figure}[ht]
\begin{center}
\includegraphics[width=0.55\textwidth,angle=0]{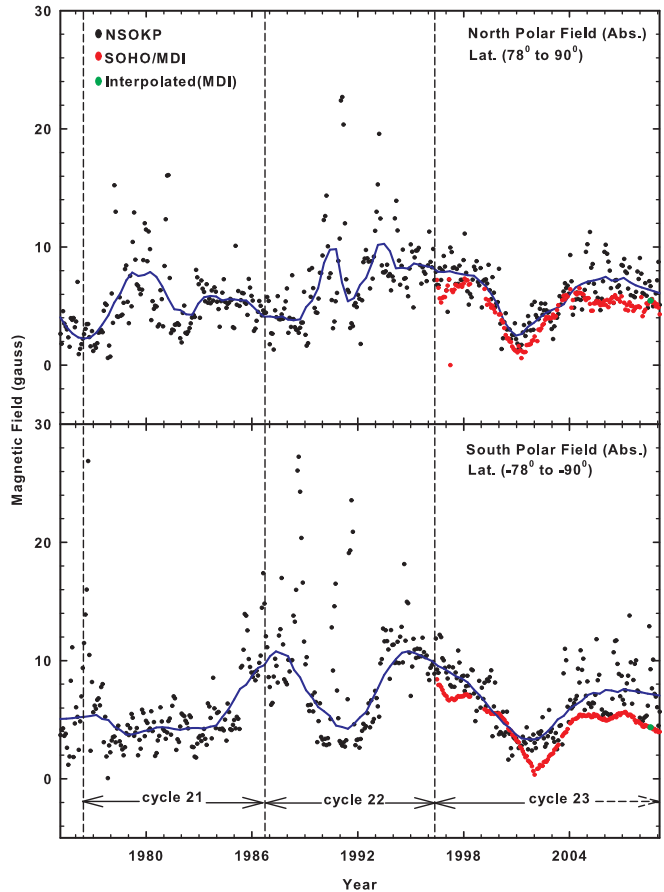}
\end{center}
\caption{{\sl\small{The two panels show the variations in the absolute value of the polar magnetic field
in the latitude range 78${^{\circ}}$--90${^{\circ}}$ for the north and south solar hemisphere as 
a function of years for the solar cycles 21--23.  The actual measurements derived from the 
NSO/Kitt peak data base is shown by filled dots while the blue solid line is a smoothed curve.  
MDI data for cycle 23 (for comparison) is shown by red dots.  The vertically oriented dashed parallel 
lines demarcate cycles 21, 22 and 23 respectively. Points where MDI data gaps were filled by interpolation 
are shown by green dots.}}}
\label{abs-fields}
\end{figure}

The two panels of Figure \ref{abs-fields} show the variation in the absolute value of the polar 
magnetic field in the latitude range 78${^{\circ}}$--90${^{\circ}}$ for the north and south solar 
hemisphere as a function of years for the solar cycles 21--23.  
%
%
During cycle 23, MDI data is shown by red dots for comparison between NSO/Kitt Peak data and MDI data.  Marked by
green dots are points 
where MDI data gaps were filled in by interpolation.  It may be noted that the MDI high 
resolution data was reduced,by averaging, to that of the NSO/Kitt Peak data for comparison.  A striking feature in 
Fig. \ref{abs-fields} is a a steep and continuous drop of the average polar field (in both NSO and MDI data) from the late declining phase of cycle 22 to the maximum of cycle 23 in both the hemispheres. A slow 
continuous drop of $\sim$13 G from 1994 to 2001 is seen in the northern hemisphere while a sharp 
continuous drop of $\sim$9 G from 1995 to 2002 is seen in the southern hemisphere.  Cycles 21 and 22 
however showed no such drop.  
The recent and extended minimum at the end of cycle 23 is also interesting in that the average polar field 
behavior showed no variation and remained steady from 2004 onwards in both the hemispheres as seen from both NSO and MDI data.  The MDI data however is systematically lower than the NSO/Kitt Peak values throughout cyclel 23.  

\section*{Polar Fields and Meridional Flow Speeds} 
      \label{S-Cycle23} 
\subsection*{Cycle 23}
\label{S-polar}

The suns meridional flow which is directed poleward at the surface of the sun, is an axis symmetric 
flow of the order of 10---20 m s${^{-1}}$.  This meridional flow is responsible for 
carrying background magnetic fields poleward from the equator and plays an important role in 
determining the strength of the solar polar field and the intensity of sunspot cycles \cite{HRi10}. There 
have been some attempts to infer surface meridional flow speeds using the helioseismic technique 
of ring-diagram analysis applied to MDI data \cite{HaH02}. However, a more recent and detailed 
study using MDI images of the line-of-sight magnetic fields \cite{HRi10} has reported 
measurements of the surface meriodional flow speed in the latitude range $\pm$75${^{\circ}}$ in 
cycle 23.  These measurements have shown a lot of variation in the meridional flow speeds in 
cycle 23.  These authors determined a meridional flow speed of 11.5 m s${^{-1}}$ at minimum of cycle 23 
in 1996~---~1997 which then dropped to 8.5 m s${^{-1}}$ at the maximum in 2000~---~2001.  It must be 
noted here
that this drop in the flow speed coincides with the steep drop in the absolute value of the magnetic 
field that is seen in both the NSO/Kitt-Peak and MDI data (see Fig. \ref{abs-fields}).  In fact, the 
MDI data in the southern hemisphere continues to drop until 2002 after which it rises again.  Between 2001 
and 2004, the measured meriodional flow speed again increased to 13.0 m s${^{-1}}$ and remained 
constant at this value thereafter.  Again, from Fig. \ref{abs-fields} we can see a corresponding 
steep increase in the absolute value of the polar field between 2001 and 2004.  After this the 
absolute value of the magnetic field remains constant in both hemispheres until 2009.  From 
Fig. \ref{abs-fields} it is clear that the polar field strength since 2004, as seen by both 
NSO/Kitt Peak and MDI has remained constant. 
%
\begin{figure}[ht]
\begin{center}
\includegraphics[width=0.55\textwidth,angle=0]{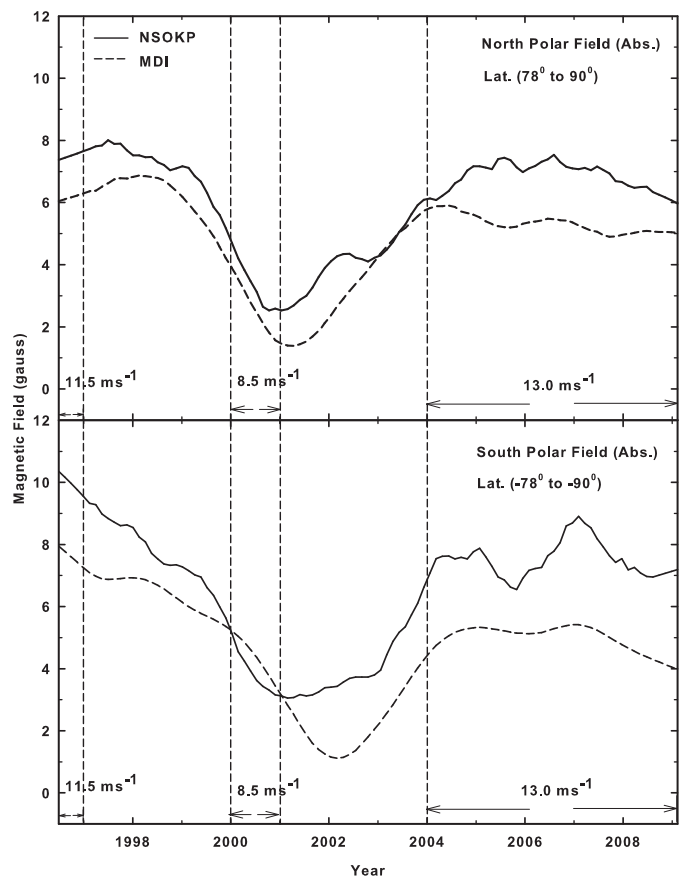}
\end{center}
\caption{{\sl\small{The two panels show a comparison between Kitt-Peak/NSO data and MDI data in cycle 23 for the 
two hemispheres.  The Northern hemisphere is shown in the upper panel while the southern hemisphere is shown 
in the lower panel.  The meridional flow speeds as reported by Hathaway \& Rightmore (2010) have been 
indicated by vertically oriented dashed parallel lines.}}}
\label{nso-mdi}
\end{figure}	    
%
Figure \ref{nso-mdi} shows a comparison between the averaged absolute magnetic
fields determined from NSO/Kitt Peak synoptic magnetogram maps and MDI synoptic magnetogram maps 
for the northern (upper panel) and southern (lower panel) hemispheres for solar cycle 23.  While 
the solid line is a smoothed curve representing the NSO data, the dashed line is a smoothed curve 
representing the MDI data.  The meridional flow speeds 
as reported by \citeauthor{HRi10}, \citeyear{HRi10} are indicated at the bottom of each panel and 
demarcated by vertically oriented dashed parallel lines.  It may be noted again that 
the MDI high resolution data (3600$\times$1080 pixels) was averaged and reduced to that of the 
NSO/Kitt Peak data (360$\times$180 pixels) for comparison. The recent and extended minimum at 
the end of cycle 23 is also interesting in that the average polar field behavior showed no 
variation and remained steady from 2004 onwards in both the hemispheres as seen from 
Figure \ref{nso-mdi}. The magnetic field of MDI data however is systematically lower than the 
NSO/Kitt Peak values throughout cycle 23 and in the southern hemisphere, the MDI data shows ta 
steep drop between 1997 -- 2002.

\subsection*{Strength of upcoming solar cycle 24}\label{S-cycle24}

In trying to predict the strength of future solar cycles, magnetic persistance or the sun's memory of the polar 
field strength at minimum in the previous cycle has been postulated as the influencing factor.  Some authors 
have used this method to try and predict the strength of upcoming of cycles through magnetic persistence or Sun's memory(\cite{ScS78}; \cite{LaF91}; \cite{SSo96}).  Later, it was postulated that magnetic persistence 
is itself governed by meridional flow speed (\cite{Hat96}; \cite{HaH02}; \cite{BAn03}) with slower 
meridional flows resulting in longer memory.  In addition, a slower meridional flow speed would also determine 
the cycle period.

Two different views were postulated to predict the strength of solar cycle 24. The first
predicted the new solar cycle 24 to be a strong cycle compared to cycle 23 \cite{DTG06}. These authors used sunspot area
of last three cycles based on flux transport dynamo model as the source term for
generating poloidal field.  The second predicted the new solar cycle 24 to be weak one \cite{CCJ07}.
These authors used only inputs from the last solar cycle since the generation of poloidal fields involved randomness and 
indeterministic behavior.

\subsection*{Discussion and Conclusions}

Though a great deal of progress has been made in our understanding of meridional flows and their 
role in determining the strength of the polar field and the amplitude of the following cycles, there seems to be a basic disagreement between flux transport dynamo models and surface flux transport models.  Fast meridional flows, in the flux transport dynamo models, produce stronger polar fields 
and a short cycle, as compared to the observations of surface flux transport which give rise to weak polar fields and a long solar cycle.  While the dynamo models have fields of one polarity centered on the sunspot latitudes, surface models have bands of opposite magnetic polarity on either side of the
sunspot latitudes.  The cause for the difference in these two models, arises mainly due to 
the above difference in the latitudinal distribution of magnetic polarities and has been adequately 
explained by \cite{HRi10}.

We have seen form our data, a large and unusual drop in the absolute value of the polar fields 
during cycle 23 compared to previous cycles (section \ref{S-Obs}) and also it's association  
with similar behavior in meridional flow speed (subsection \ref{S-polar}). In addition, \cite{HRi10} 
has shown that the meridional flow generally shows a decrease from minimum to maximum. This has also 
been reported for previous solar cycles \cite{KHH93} with less certainty and poorer time resolution than 
for cycle 23 where, high resolution MDI data is available. The minimum leading up to the start 
of cycle 24 has been very deep and prolonged and we have seen that the corresponding polar fields 
have been at their lowest compared to the other cycles 21 and 22.  We believe that 
the memory of this very weak polar field in cycle 23 is the cause of the extended solar minimum 
we have witnessed. Since the sun's memory depends on polar field strength we believe that the 
upcoming cycle 24 will be a much weaker cycle compared to cycle 23 which is in accordance with 
the later prediction proposed by \cite{CCJ07}.

We have seen above that the absolute value of the polar magnetic field in the latitude bin 
78${^{\circ}}$--90${^{\circ}}$ virtually mirrors the change in the meridional flow speeds 
in cycle 23, with meridional flow speeds dropping when the absolute value of the field drops. 
Unlike cycles 21 and 22, the absolute values of the polar magnetic fields in cycle 23  showed 
similar trends in both hemispheres.  Since high resolution MDI data is available only from 1996, 
it is not be possible to compute meridional flow speeds during cycles 21 and 22.  However, it 
may be noted that in both cycles 21 and 23, the behaviour of the absolute value of the polar 
fields is different in each hemisphere.  If we assume that the correlation between the absolute 
value of the polar fields and the meridional flow velocities holds good for all solar 
cycles, then this would imply that the field reversals occured at different times in the 
two hemispheres in both cycles 21, 22 and 23.  

The study of polar magnetic fields can thus provide a good clues on the nature of the meridional 
flows in each hemisphere.  Such inputs would be important in predicting the behaviour of the 
future solar cycles.  In addition to such measurements other inputs from radio measurements of 
circular polarization that are sensitive to the lineof-sight component of the coronal magnetic 
field or high resolution high dynamic range quiet sun radio imaging \cite{MeS06} of the sun at 
low frequencies will also be very useful.

\section*{{\bf{Acknowlwdgements}}}

{\sl{The authors would thank the free use data policy of the national Solar Observatory and 
acknowledge the MDI consortium for providing the data in the public domain via the world
wide web. SoHO is a project of international collaboration between ESA and NASA.}}


\newpage
\begin{thebibliography}{27}
\ifx \bisbn   \undefined \def \bi
sbn  #1{ISBN #1}\fi
\ifx \binits  \undefined \def \binits#1{#1}\fi
\ifx \bauthor  \undefined \def \bauthor#1{#1}\fi
\ifx \batitle  \undefined \def \batitle#1{#1}\fi
\ifx \bjtitle  \undefined \def \bjtitle#1{\textit{#1}}\fi
\ifx \bvolume  \undefined \def \bvolume#1{\textbf{#1}}\fi
\ifx \byear  \undefined \def \byear#1{#1}\fi
\ifx \bissue  \undefined \def \bissue#1{#1}\fi
\ifx \bfpage  \undefined \def \bfpage#1{#1}\fi
\ifx \blpage  \undefined \def \blpage #1{#1}\fi
\ifx \burl  \undefined \def \burl#1{\textsf{#1}}\fi
\ifx \href  \undefined \def \href#1#2{\textsf{#2}}\fi
\ifx \doiurl  \undefined \def
  \doiurl#1{\href{http://dx.doi.org/#1}{\textsf{#1}}}\fi
\ifx \betal  \undefined \def \betal{\textit{et al.}}\fi
\ifx \binstitute  \undefined \def \binstitute#1{#1}\fi
\ifx \bctitle  \undefined \def \bctitle#1{#1}\fi
\ifx \beditor  \undefined \def \beditor#1{#1}\fi
\ifx \bpublisher  \undefined \def \bpublisher#1{#1}\fi
\ifx \bbtitle  \undefined \def \bbtitle#1{\textit{#1}}\fi
\ifx \bedition  \undefined \def \bedition#1{#1}\fi
\ifx \bseriesno  \undefined \def \bseriesno#1{\textbf{#1}}\fi
\ifx \blocation  \undefined \def \blocation#1{#1}\fi
\ifx \bsertitle  \undefined \def \bsertitle#1{\textit{#1}}\fi
\ifx \bsnm \undefined \def \bsnm#1{#1}\fi
\ifx \bsuffix \undefined \def \bsuffix#1{#1}\fi
\ifx \bparticle \undefined \def \bparticle#1{#1}\fi
\ifx \barticle \undefined \def \barticle{}\fi
\ifx \botherref \undefined \def \botherref{}\fi
\ifx \url \undefined \def \url#1{\textsf{#1}}\fi
\ifx \bchapter \undefined \def \bchapter{}\fi
\ifx \bbook \undefined \def \bbook{}\fi
\ifx \bcomment \undefined \def \bcomment#1{#1}\fi
\ifx \oauthor \undefined \def \oauthor#1{#1}\fi
\ifx \citeauthoryear \undefined \def \citeauthoryear#1{#1}\fi
\def \endbibitem {}

\bibitem[\protect\citeauthoryear{{Babcock}}{1959}]{Bab59}
\begin{barticle}
\bauthor{\bsnm{{Babcock}}, \binits{H.D.}}:
\byear{1959},
\batitle{{The Sun's Polar Magnetic Field.}}
\bjtitle{\apj}
\bvolume{130},
\bfpage{364}.
doi:\doiurl{10.1086/146726}.
\end{barticle}
\endbibitem

\bibitem[\protect\citeauthoryear{{Babcock} and {Babcock}}{1955}]{BBa55}
\begin{barticle}
\bauthor{\bsnm{{Babcock}}, \binits{H.W.}}, \bauthor{\bsnm{{Babcock}},
  \binits{H.D.}}:
\byear{1955},
\batitle{{The Sun's Magnetic Field, 1952-1954.}}
\bjtitle{\apj}
\bvolume{121},
\bfpage{349}.
doi:\doiurl{10.1086/145994}.
\end{barticle}
\endbibitem

\bibitem[\protect\citeauthoryear{{Basu} and {Antia}}{2003}]{BAn03}
\begin{barticle}
\bauthor{\bsnm{{Basu}}, \binits{S.}}, \bauthor{\bsnm{{Antia}}, \binits{H.M.}}:
\byear{2003},
\batitle{{Changes in Solar Dynamics from 1995 to 2002}}.
\bjtitle{\apj}
\bvolume{585},
\bfpage{553}\,--\,\blpage{565}.
doi:\doiurl{10.1086/346020}.
\end{barticle}
\endbibitem

\bibitem[\protect\citeauthoryear{{Benevolenskaya}, {Kosovichev}, and
  {Scherrer}}{2001}]{BKS01}
\begin{barticle}
\bauthor{\bsnm{{Benevolenskaya}}, \binits{E.E.}}, \bauthor{\bsnm{{Kosovichev}},
  \binits{A.G.}}, \bauthor{\bsnm{{Scherrer}}, \binits{P.H.}}:
\byear{2001},
\batitle{{Detection of High-Latitude Waves of Solar Coronal Activity in
  Extreme-Ultraviolet Data from the Solar and Heliospheric Observatory EUV
  Imaging Telescope}}.
\bjtitle{\apjl}
\bvolume{554},
\bfpage{L107}\,--\,\blpage{L110}.
doi:\doiurl{10.1086/320925}.
\end{barticle}
\endbibitem

\bibitem[\protect\citeauthoryear{{Benevolenskaya}, {Kosovichev}, and
  {Scherrer}}{2002}]{BKS02}
\begin{botherref}
\oauthor{\bsnm{{Benevolenskaya}}, \binits{E.E.}}, \oauthor{\bsnm{{Kosovichev}},
  \binits{A.G.}}, \oauthor{\bsnm{{Scherrer}}, \binits{P.H.}}:
2002,
{Evolution of Large-scale Coronal Structure with the Solar Cycle from EUV
  Data}.
In: {F.~Favata \& J.~J.~Drake} (ed.)
\textit{Stellar Coronae in the Chandra and XMM-NEWTON Era},
\textit{Astronomical Society of the Pacific Conference Series}
\textbf{277},
419.
\end{botherref}
\endbibitem

\bibitem[\protect\citeauthoryear{{Choudhuri}, {Chatterjee}, and
  {Jiang}}{2007}]{CCJ07}
\begin{barticle}
\bauthor{\bsnm{{Choudhuri}}, \binits{A.R.}}, \bauthor{\bsnm{{Chatterjee}},
  \binits{P.}}, \bauthor{\bsnm{{Jiang}}, \binits{J.}}:
\byear{2007},
\batitle{{Predicting Solar Cycle 24 With a Solar Dynamo Model}}.
\bjtitle{Physical Review Letters}
\bvolume{98}(\bissue{13}),
\bfpage{131103}.
doi:\doiurl{10.1103/PhysRevLett.98.131103}.
\end{barticle}
\endbibitem

\bibitem[\protect\citeauthoryear{{Choudhuri}, {Schussler}, and
  {Dikpati}}{1995}]{CSD95}
\begin{barticle}
\bauthor{\bsnm{{Choudhuri}}, \binits{A.R.}}, \bauthor{\bsnm{{Schussler}},
  \binits{M.}}, \bauthor{\bsnm{{Dikpati}}, \binits{M.}}:
\byear{1995},
\batitle{{The solar dynamo with meridional circulation.}}
\bjtitle{\aap}
\bvolume{303},
\bfpage{L29}.
\end{barticle}
\endbibitem

\bibitem[\protect\citeauthoryear{{Dikpati} and {Charbonneau}}{1999}]{DCh99}
\begin{barticle}
\bauthor{\bsnm{{Dikpati}}, \binits{M.}}, \bauthor{\bsnm{{Charbonneau}},
  \binits{P.}}:
\byear{1999},
\batitle{{A Babcock-Leighton Flux Transport Dynamo with Solar-like Differential
  Rotation}}.
\bjtitle{\apj}
\bvolume{518},
\bfpage{508}\,--\,\blpage{520}.
doi:\doiurl{10.1086/307269}.
\end{barticle}
\endbibitem

\bibitem[\protect\citeauthoryear{{Dikpati}, {de Toma}, and
  {Gilman}}{2006}]{DTG06}
\begin{barticle}
\bauthor{\bsnm{{Dikpati}}, \binits{M.}}, \bauthor{\bsnm{{de Toma}},
  \binits{G.}}, \bauthor{\bsnm{{Gilman}}, \binits{P.A.}}:
\byear{2006},
\batitle{{Predicting the strength of solar cycle 24 using a flux-transport
  dynamo-based tool}}.
\bjtitle{\grl}
\bvolume{33},
\bfpage{5102}.
doi:\doiurl{10.1029/2005GL025221}.
\end{barticle}
\endbibitem

\bibitem[\protect\citeauthoryear{{Dikpati} \textit{et~al.}}{2004}]{DiG04}
\begin{barticle}
\bauthor{\bsnm{{Dikpati}}, \binits{M.}}, \bauthor{\bsnm{{de Toma}},
  \binits{G.}}, \bauthor{\bsnm{{Gilman}}, \binits{P.A.}},
  \bauthor{\bsnm{{Arge}}, \binits{C.N.}}, \bauthor{\bsnm{{White}},
  \binits{O.R.}}:
\byear{2004},
\batitle{{Diagnostics of Polar Field Reversal in Solar Cycle 23 Using a Flux
  Transport Dynamo Model}}.
\bjtitle{\apj}
\bvolume{601},
\bfpage{1136}\,--\,\blpage{1151}.
doi:\doiurl{10.1086/380508}.
\end{barticle}
\endbibitem

\bibitem[\protect\citeauthoryear{{Domingo}, {Fleck}, and
  {Poland}}{1995}]{DFP95}
\begin{barticle}
\bauthor{\bsnm{{Domingo}}, \binits{V.}}, \bauthor{\bsnm{{Fleck}}, \binits{B.}},
  \bauthor{\bsnm{{Poland}}, \binits{A.I.}}:
\byear{1995},
\batitle{{The SOHO Mission: an Overview}}.
\bjtitle{\solphys}
\bvolume{162},
\bfpage{1}\,--\,\blpage{2}.
doi:\doiurl{10.1007/BF00733425}.
\end{barticle}
\endbibitem

\bibitem[\protect\citeauthoryear{{Duvall}}{1979}]{Duv79}
\begin{barticle}
\bauthor{\bsnm{{Duvall}}, \binits{T.L.} \bsuffix{Jr.}}:
\byear{1979},
\batitle{{Large-scale solar velocity fields}}.
\bjtitle{\solphys}
\bvolume{63},
\bfpage{3}\,--\,\blpage{15}.
doi:\doiurl{10.1007/BF00155690}.
\end{barticle}
\endbibitem

\bibitem[\protect\citeauthoryear{{Fox}, {McIntosh}, and {Wilson}}{1998}]{FMW98}
\begin{barticle}
\bauthor{\bsnm{{Fox}}, \binits{P.}}, \bauthor{\bsnm{{McIntosh}}, \binits{P.}},
  \bauthor{\bsnm{{Wilson}}, \binits{P.R.}}:
\byear{1998},
\batitle{{Coronal Holes and the Polar Field Reversals}}.
\bjtitle{\solphys}
\bvolume{177},
\bfpage{375}\,--\,\blpage{393}.
\end{barticle}
\endbibitem

\bibitem[\protect\citeauthoryear{{Gopalswamy} \textit{et~al.}}{2003}]{GoL03}
\begin{barticle}
\bauthor{\bsnm{{Gopalswamy}}, \binits{N.}}, \bauthor{\bsnm{{Lara}},
  \binits{A.}}, \bauthor{\bsnm{{Yashiro}}, \binits{S.}},
  \bauthor{\bsnm{{Howard}}, \binits{R.A.}}:
\byear{2003},
\batitle{{Coronal Mass Ejections and Solar Polarity Reversal}}.
\bjtitle{\apjl}
\bvolume{598},
\bfpage{L63}\,--\,\blpage{L66}.
doi:\doiurl{10.1086/380430}.
\end{barticle}
\endbibitem

\bibitem[\protect\citeauthoryear{{Haber} \textit{et~al.}}{2002}]{HaH02}
\begin{barticle}
\bauthor{\bsnm{{Haber}}, \binits{D.A.}}, \bauthor{\bsnm{{Hindman}},
  \binits{B.W.}}, \bauthor{\bsnm{{Toomre}}, \binits{J.}},
  \bauthor{\bsnm{{Bogart}}, \binits{R.S.}}, \bauthor{\bsnm{{Larsen}},
  \binits{R.M.}}, \bauthor{\bsnm{{Hill}}, \binits{F.}}:
\byear{2002},
\batitle{{Evolving Submerged Meridional Circulation Cells within the Upper
  Convection Zone Revealed by Ring-Diagram Analysis}}.
\bjtitle{\apj}
\bvolume{570},
\bfpage{855}\,--\,\blpage{864}.
doi:\doiurl{10.1086/339631}.
\end{barticle}
\endbibitem

\bibitem[\protect\citeauthoryear{{Hathaway}}{1996}]{Hat96}
\begin{barticle}
\bauthor{\bsnm{{Hathaway}}, \binits{D.H.}}:
\byear{1996},
\batitle{{Doppler Measurements of the Sun's Meridional Flow}}.
\bjtitle{\apj}
\bvolume{460},
\bfpage{1027}.
doi:\doiurl{10.1086/177029}.
\end{barticle}
\endbibitem

\bibitem[\protect\citeauthoryear{{Hathaway} and {Rightmire}}{2010}]{HRi10}
\begin{barticle}
\bauthor{\bsnm{{Hathaway}}, \binits{D.H.}}, \bauthor{\bsnm{{Rightmire}},
  \binits{L.}}:
\byear{2010},
\batitle{{Variations in the Suns Meridional Flow over a Solar Cycle}}.
\bjtitle{Science}
\bvolume{327},
\bfpage{1350}.
doi:\doiurl{10.1126/science.1181990}.
\end{barticle}
\endbibitem

\bibitem[\protect\citeauthoryear{{Hathaway} \textit{et~al.}}{1996}]{HaG96}
\begin{barticle}
\bauthor{\bsnm{{Hathaway}}, \binits{D.H.}}, \bauthor{\bsnm{{Gilman}},
  \binits{P.A.}}, \bauthor{\bsnm{{Harvey}}, \binits{J.W.}},
  \bauthor{\bsnm{{Hill}}, \binits{F.}}, \bauthor{\bsnm{{Howard}},
  \binits{R.F.}}, \bauthor{\bsnm{{Jones}}, \binits{H.P.}},
  \bauthor{\bsnm{{Kasher}}, \binits{J.C.}}, \bauthor{\bsnm{{Leibacher}},
  \binits{J.W.}}, \bauthor{\bsnm{{Pintar}}, \binits{J.A.}},
  \bauthor{\bsnm{{Simon}}, \binits{G.W.}}:
\byear{1996},
\batitle{{GONG Observations of Solar Surface Flows}}.
\bjtitle{Science}
\bvolume{272},
\bfpage{1306}\,--\,\blpage{1309}.
doi:\doiurl{10.1126/science.272.5266.1306}.
\end{barticle}
\endbibitem

\bibitem[\protect\citeauthoryear{{Komm}, {Howard}, and {Harvey}}{1993}]{KHH93}
\begin{barticle}
\bauthor{\bsnm{{Komm}}, \binits{R.W.}}, \bauthor{\bsnm{{Howard}},
  \binits{R.F.}}, \bauthor{\bsnm{{Harvey}}, \binits{J.W.}}:
\byear{1993},
\batitle{{Meridional Flow of Small Photospheric Magnetic Features}}.
\bjtitle{\solphys}
\bvolume{147},
\bfpage{207}\,--\,\blpage{223}.
doi:\doiurl{10.1007/BF00690713}.
\end{barticle}
\endbibitem

\bibitem[\protect\citeauthoryear{{Layden} \textit{et~al.}}{1991}]{LaF91}
\begin{barticle}
\bauthor{\bsnm{{Layden}}, \binits{A.C.}}, \bauthor{\bsnm{{Fox}},
  \binits{P.A.}}, \bauthor{\bsnm{{Howard}}, \binits{J.M.}},
  \bauthor{\bsnm{{Sarajedini}}, \binits{A.}}, \bauthor{\bsnm{{Schatten}},
  \binits{K.H.}}:
\byear{1991},
\batitle{{Dynamo-based scheme for forecasting the magnitude of solar activity
  cycles}}.
\bjtitle{\solphys}
\bvolume{132},
\bfpage{1}\,--\,\blpage{40}.
doi:\doiurl{10.1007/BF00159127}.
\end{barticle}
\endbibitem

\bibitem[\protect\citeauthoryear{{Mercier} \textit{et~al.}}{2006}]{MeS06}
\begin{barticle}
\bauthor{\bsnm{{Mercier}}, \binits{C.}}, \bauthor{\bsnm{{Subramanian}},
  \binits{P.}}, \bauthor{\bsnm{{Kerdraon}}, \binits{A.}},
  \bauthor{\bsnm{{Pick}}, \binits{M.}}, \bauthor{\bsnm{{Ananthakrishnan}},
  \binits{S.}}, \bauthor{\bsnm{{Janardhan}}, \binits{P.}}:
\byear{2006},
\batitle{{Combining visibilities from the giant meterwave radio telescope and
  the Nancay radio heliograph. High dynamic range snapshot images of the solar
  corona at 327 MHz}}.
\bjtitle{\aap}
\bvolume{447},
\bfpage{1189}\,--\,\blpage{1201}.
doi:\doiurl{10.1051/0004-6361:20053621}.
\end{barticle}
\endbibitem

\bibitem[\protect\citeauthoryear{{Parker}}{1955a}]{Par55a}
\begin{barticle}
\bauthor{\bsnm{{Parker}}, \binits{E.N.}}:
\byear{1955}a,
\batitle{{Hydromagnetic Dynamo Models.}}
\bjtitle{\apj}
\bvolume{122},
\bfpage{293}.
doi:\doiurl{10.1086/146087}.
\end{barticle}
\endbibitem

\bibitem[\protect\citeauthoryear{{Parker}}{1955b}]{Par55b}
\begin{barticle}
\bauthor{\bsnm{{Parker}}, \binits{E.N.}}:
\byear{1955}b,
\batitle{{The Formation of Sunspots from the Solar Toroidal Field.}}
\bjtitle{\apj}
\bvolume{121},
\bfpage{491}.
doi:\doiurl{10.1086/146010}.
\end{barticle}
\endbibitem

\bibitem[\protect\citeauthoryear{{Schatten} and {Sofia}}{1996}]{SSo96}
\begin{botherref}
\oauthor{\bsnm{{Schatten}}, \binits{K.}}, \oauthor{\bsnm{{Sofia}},
  \binits{S.}}:
1996,
{Forecasting Solar Activity and Cycle 23 Outlook}.
In: \textit{Bulletin of the American Astronomical Society},
\textit{Bulletin of the American Astronomical Society}
\textbf{28},
1347.
\end{botherref}
\endbibitem

\bibitem[\protect\citeauthoryear{{Schatten} \textit{et~al.}}{1978}]{ScS78}
\begin{barticle}
\bauthor{\bsnm{{Schatten}}, \binits{K.H.}}, \bauthor{\bsnm{{Scherrer}},
  \binits{P.H.}}, \bauthor{\bsnm{{Svalgaard}}, \binits{L.}},
  \bauthor{\bsnm{{Wilcox}}, \binits{J.M.}}:
\byear{1978},
\batitle{{Using dynamo theory to predict the sunspot number during solar cycle
  21}}.
\bjtitle{\grl}
\bvolume{5},
\bfpage{411}\,--\,\blpage{414}.
doi:\doiurl{10.1029/GL005i005p00411}.
\end{barticle}
\endbibitem

\bibitem[\protect\citeauthoryear{{Scherrer} \textit{et~al.}}{1995}]{ScB95}
\begin{barticle}
\bauthor{\bsnm{{Scherrer}}, \binits{P.H.}}, \bauthor{\bsnm{{Bogart}},
  \binits{R.S.}}, \bauthor{\bsnm{{Bush}}, \binits{R.I.}},
  \bauthor{\bsnm{{Hoeksema}}, \binits{J.T.}}, \bauthor{\bsnm{{Kosovichev}},
  \binits{A.G.}}, \bauthor{\bsnm{{Schou}}, \binits{J.}},
  \bauthor{\bsnm{{Rosenberg}}, \binits{W.}}, \bauthor{\bsnm{{Springer}},
  \binits{L.}}, \bauthor{\bsnm{{Tarbell}}, \binits{T.D.}},
  \bauthor{\bsnm{{Title}}, \binits{A.}}, \bauthor{\bsnm{{Wolfson}},
  \binits{C.J.}}, \bauthor{\bsnm{{Zayer}}, \binits{I.}}, \bauthor{\bsnm{{MDI
  Engineering Team}}}:
\byear{1995},
\batitle{{The Solar Oscillations Investigation - Michelson Doppler Imager}}.
\bjtitle{\solphys}
\bvolume{162},
\bfpage{129}\,--\,\blpage{188}.
doi:\doiurl{10.1007/BF00733429}.
\end{barticle}
\endbibitem

\bibitem[\protect\citeauthoryear{{Steenbeck} and {Krause}}{1969}]{SKr69}
\begin{barticle}
\bauthor{\bsnm{{Steenbeck}}, \binits{M.}}, \bauthor{\bsnm{{Krause}},
  \binits{F.}}:
\byear{1969},
\batitle{{On the Dynamo Theory of Stellar and Planetary Magnetic Fields. I. AC
  Dynamos of Solar Type}}.
\bjtitle{Astronomische Nachrichten}
\bvolume{291},
\bfpage{49}\,--\,\blpage{84}.
\end{barticle}
\endbibitem

\bibitem[\protect\citeauthoryear{{Wang}, {Sheeley}, and {Nash}}{1991}]{WSN91}
\begin{barticle}
\bauthor{\bsnm{{Wang}}, \binits{Y.}}, \bauthor{\bsnm{{Sheeley}}, \binits{N.R.}
  \bsuffix{Jr.}}, \bauthor{\bsnm{{Nash}}, \binits{A.G.}}:
\byear{1991},
\batitle{{A new solar cycle model including meridional circulation}}.
\bjtitle{\apj}
\bvolume{383},
\bfpage{431}\,--\,\blpage{442}.
doi:\doiurl{10.1086/170800}.
\end{barticle}
\endbibitem

\end{thebibliography}
\end{document}